\documentstyle[12pt,A4,epsfig]{article}

\newcommand{\be}{\begin{equation}}
\newcommand{\ee}{\end{equation}}
\newcommand{\bea}{\begin{eqnarray}}
\newcommand{\eea}{\end{eqnarray}}
\begin{document}
\begin{titlepage}
\begin{flushright}
DO-TH-94/06\\
March 1994
\end{flushright}

\vspace{20mm}
\begin{center}
{\Large \bf
The Fermion Determinant in External Scalar Fields: Exact Results
and Large Mass Approximation
}
\vspace{10mm}

{\large  J. Baacke\footnote{e-mail~~  baacke@het.physik.uni-dortmund.de},
H. So
\footnote{On leave of absence from Dept. of Physics,
Niigata University, Niigata, Japan} 
\footnote{e-mail~~  so@het.physik.uni-dortmund.de}
and A. S\"urig\footnote{e-mail~~  suerig@het.physik.uni-dortmund.de}} \\
\vspace{15mm}

{\large Institut f\"ur Physik, Universit\"at Dortmund} \\
{\large D - 44221 Dortmund , Germany}
\vspace{25mm}

\bf{Abstract}
\end{center}
We present results for the fermion determinant in an external
Yukawa-coupled scalar field, a quantity relevant to instanton
and sphaleron transition rates as well as bubble nucleation occuring
in the electroweak phase transition when the scalar
field is identified with the Higgs field. We calculate the determinant
exactly and find an approximation (essentialy of the gradient expansion
type) that reproduces the exact results very well if the fermion mass $m_F$
becomes larger than the inverse of the size  $R$ of the background field
configuration. The results are given for 2, 3 and 4 Euclidean
dimensions.
\end{titlepage}
\setcounter{footnote}{0}

\section{Introduction}
\par

The fermion determinant in an external Higgs field is a quantity
that enters semiclassical rates \cite{La,Aff}
for sphaleron \cite{Ma,KliMa,ArMcL} 
and instanton \cite{tHoo} transitions as
well as for the nucleation of  bubbles or droplets
\cite{Co1,CaCo} in first order phase transitions,
as possibly the electroweak phase transition occuring in the
early universe. The classical action is usually minimized by
spherically symmetric field configurations,
the mass term produced by the Higgs field 
is then given by a radial profile $\phi(x)=f(r)$. We will consider
here field configurations that approach the broken Higgs
phase as $r \to \infty$. 

The exact calculation of such determinants has to be
performed numerically since usually even the background field
configuration is only given as a numerical profile. A very
efficient method based on a theorem on functional determinants
\cite{theorem} has been developed recently by V. G. Kiselev and one of
the authors (J.B.) \cite{BaaKi}. We will apply this method here
for evaluating the fermion determinant exactly in 2, 3 and 4
Euclidean dimensions.

It is certainly useful to dispose of analytical approximations
to the fermion determinant, especially if the back reaction of
this determinant on the dynamics of the Higgs field is being
considered in the form of an effective action. It is our main
purpose here to present such an approximation and, using our
exact results, to demonstrate that it approximates the
fermion determinant very well for sufficiently large fermion masses.
This covers indeed the most relevant case since the effective action
induced by fermion background fields behaves essentially as
$m_F^D$ where $D$ is the Euclidean dimension and so is most relevant
for heavy fermions.

The paper is organized as follows:
In section 2 we recall some general relations and definitions.
In section 3 we present the spherical decomposition
of the relevant Dirac equation in $D=2,3$ and $4$ and discuss the
numerical evaluation of the determinant. In section 4 we
develop our approximation. The numerical and approximate analytical
results are presented and discussed in section 5. Details
on the Dirac operators and the partial wave decomposition are
given in the Appendix.
\setcounter{equation}{0}
\section{Basic Relations}
The Dirac operator for a fermion with Yukawa coupling $g_Y$ to an
external scalar field takes, in Euclidean space, the form
\be
{\cal D} = \gamma_\mu \partial_\mu - m_F \phi (x)
={\cal D}^{(0)} - m_F(\phi(x)-1)
\ee
where we have normalized the ``Higgs'' field to its vacuum
expectation value $v$ so that $m_F = g_Y v$ and where
${\cal D}^{(0)}$ is the free Dirac operator in the ``broken
Higgs phase'' where $\phi(x) = 1$.
The gamma matrices are
hermitean and satisfy $\gamma_\mu \gamma_\nu+
\gamma_\nu\gamma_\mu= 2 \delta_{\mu\nu}$.
We also define the positive definite operator
\be
{\cal M} = {\cal D}^\dagger{\cal D} =
-\partial^2+ m_F^2 + {\cal V}(x)
\label{Mdef}\ee
where the potential or vertex matrix ${\cal V}$ takes the form
\be
{\cal V}(x) = m_F^2(\phi(x)^2-1) + m_F \gamma_\mu\partial_\mu \phi(x)
\ee
We can evaluate the determinant of the Dirac operator, normalized to the
free Dirac operator in the form
\bea
&\ln \det ({\cal D}/{\cal D}^{(0)})= 
\frac{1}{2} \ln {\cal J}&
\\ \nonumber
&{\cal J} = \det \left(
\displaystyle{\frac{-\partial^2+m_F^2+V(x)}{-\partial^2+m_F^2}}
\right)&
\eea
Furthermore the fermionic fluctuation determinant 
is related to the corresponding one-loop
effective action by the relation
\be
S_{eff}[\phi] = -\frac1{2} \ln {\cal J}
\label{Sdef}
 \ee

The case $D=2$ could be relevant for lattice simulations
of bubble nucleation at low temperature.
The case $D=3$ will be considered
with hindsight of an application as high temperature limit of the
four dimensional theory.
The four dimensional case which would be relevant to bubble
formation at low temperatures (instanton type transition).
The details on the $\gamma$ matrices are given in Appendix A.

\setcounter{equation}{0}
\section{Numerical Computation of the Fluctuation Determinant}
\par    \label{determinant}

For spherically symmetric background fields $\phi(x)=f(r)$
the determinant decomposes
into contributions of partial waves in the form
\be
\ln {\cal J} = \sum d_\lambda(D) \ln J_\lambda
\ee
where $\lambda$ denotes a coupled channel of partial waves,
$J_\lambda$ is the partial wave determinant and $d_\lambda(D)$ 
is the degeneracy which depends on the dimension $D$.
$J_\lambda$ is related to the partial wave reduction of the operator
${\cal M}$ which takes the general form
\bea
{\bf M}_\lambda&=&{\bf M}_\lambda^{(0)} + {\bf V}(r)
\nonumber \\ {\bf M}_\lambda^{(0)} &=&
-\frac{d^2}{dr^2} - \frac{D-1}{r}\frac{d}{dr}
+\frac{{\bf L}_\lambda^2}{r^2} +m_F^2
\eea
via
\be
J_\lambda =  \det \left(\frac{{\bf M}_\lambda}{{\bf M}_\lambda^{(0)}}
\right)
\ee
The degeneracies $d_\lambda$, and the  matrices
${\bf L}^2_\lambda$ specifying the centrifugal barrier and
${\bf V(r)}$ are given in Appendix A.

For the numerical computation of the determinant  $J_\lambda$
we use the method described in Ref \cite{BaaKi}, generalized
to a coupled channel system. It is based on a theorem on functional
determinants \cite{theorem} which we recall below. The theorem has
been used for a similar purpose for the case of massless
fermions \cite{AvdVe}. In the following
we consider the contribution of one channel $\lambda$.

Let $\psi_{i\alpha}(r)$ be a fundamental system
solutions of the differential equation
\be
{\bf M}_\lambda \psi_{i\alpha} =0
\ee
and let $\psi_{i\alpha}^{(0)}$ be their free counterparts, both regular
at $r=0$. Since there are as
many independent solutions labelled by $\alpha$ as there are components
labelled by $i$ these fundamental systems form quadratic matrices
denoted by $\Psi = \{\psi_{i\alpha}\}$ and 
$\Psi^{(0)} = \{\psi^{(0)}_{i\alpha}\}$ 
respectively. Let furthermore
the solutions be normalized such that
\be
\lim_{r \to 0} \Psi(r)
(\Psi^{(0)}(r))^{-1} ={\bf 1}.
\ee

Then according to the theorem on functional determinants \cite{theorem}
the partial wave determinant is obtained as
\be
\det \left(\frac{{\bf M}_\lambda}{{\bf M}^{(0)}_\lambda}\right)=
\lim_{r \to \infty}
\frac{\det \Psi(r)}{\det \Psi^{(0)}(r)}
\ee

In the cases considered here --- the Dirac operator with a scalar mass
term in two, three and four dimensions --- the partial wave reduction
results in simple $(2 \times 2)$ systems. The ``free'' differential
operator ${\bf M}^{(0)}_\lambda$ is diagonal and the solutions are modified
Bessel functions which be denote as $b_\lambda$ for the upper component;
their detailed form is given in Appendix A
(see Eq. (\ref{bdef})). The solution for the
lower component ist $b_{\lambda+1}$ in all cases considered here.

The solutions of the full operator ${\bf M}_\lambda$ are now written as
\be
\psi_{i\alpha}(r) = b_{\lambda+i-1}(r) (\delta_{i\alpha} + h_{i\alpha}(r))
\ee
where the boundary condition at $r=0$ requires the functions 
$h_{i\alpha}(r)$ to vanish there. Since the Bessel functions
can be taken out of the rows of the matrix ${\bf \psi}$ when
forming the determinant they cancel in $J_\lambda$ and we have
simply
\be
J_\lambda = \lim_{r \to \infty}
\det ({\bf 1}+{\bf h}(r)).
\ee 

As in \cite{BaaKi} we write directly a differential equation
for the functions $h_{i\alpha}$. It takes the form
\be
{\bf \Delta}_\lambda {\bf h} = {\bf V}_\lambda ({\bf 1} + {\bf h})
\ee
where the differential operator ${\bf \Delta}_\lambda$
and the modified potential ${\bf V}_\lambda$ are
defined in Eqs. (\ref{hequation1}-\ref{hequation3}).
This differential equation is inhomogeneous and ${\bf h}$ may be
expanded w.r.t. orders in the potential ${\bf V}$. Denoting the
contribution of order $k$ in the potential as ${\bf h}^{(k)}$ we
introduce also the contributions summed from $k$ to $\infty$ as
\be
{\bf h}^{\overline{(k)}} = \sum_{l=k}^\infty {\bf h}^{(l)}
\ee
so that ${\bf h}={\bf h}^{\overline{(1)}}$. The differential equation
can then be used to get the functions ${\bf h}$ of various orders
recursively via
\bea
{\bf \Delta}_n {\bf h}^{(k+1)}&=& {\bf V}_n {\bf h}^{(k)}
\nonumber \\
{\bf \Delta}_n {\bf h}^{\overline{(k+1)}}&=&
{\bf V}_n {\bf h}^{\overline{(k)}}
\eea
where further ${\bf h}^{(0)} \equiv {\bf 1}$.
Using these contributions of specific order in ${\bf V}$ it is
easy to get rid of the leading, divergent contributions to
$\ln J_\lambda$ and therefore to $\ln {\cal J}$. This procedure
has been discussed in \cite{BaaKi} and can be performed
analogously for a coupled system: In $D=2$ and $D=3$ the contribution
of order $k=1$ has to be omitted, in $D=4$ the orders $k=1$ and $2$;
the finite parts of $ln J_\lambda$ are then given by
\be
(\ln J_\lambda)^{\overline{(2)}}=
\lim_{r \to \infty} \left({\rm Tr}[ \ln ({\bf 1}+{\bf h})
-{\bf h}] + {\rm Tr}( {\bf h}^{\overline{(2)}})\right)
\ee
and
\bea
(\ln J_\lambda)^{\overline{(3)}}&=&
\lim_{r \to \infty} \left( {\rm Tr}[\ln({\bf 1}+{\bf h})
-{\bf h}+ \frac1{2} {\bf h}^2] \right. 
\nonumber \\
&+&\left. {\rm Tr}( {\bf h}^{\overline{(3)}})
-\frac1{2} {\rm Tr} ({\bf h}^{\overline{(2)}}
({\bf h}^{\overline{(1)}}+{\bf h}^{(1)})) \right)
\eea
The matrices in square brackets can be diagonalized in terms
of the
eigenvalues of ${\bf h}$ and reduce then to ``subtracted
logarithms''  which can be evaluated with arbitrary precision.
The remaining terms are separately of order $\overline{(2)}$ and
of order$\overline{(3)}$ in the first and second equation
respectively. These expressions do not contain therefore
any dangereous small differences of large numbers.
The total fluctuation determinants are denoted correspondingly as
$(\ln {\cal J})^{\overline{(2)}}$
and $(\ln {\cal J})^{\overline{(3)}}$.

The divergent contributions have to be regularized and
renormalized and this depends on the specific model under consideration.
These contributions can easily be calculated analytically, at most
one needs the Fourier transform of the profile (see e.g. \cite{BaaKi}
where this step was included). Here we are interested in the comparison
with the large $m_F$ expansion and for this purpose it is sufficient to
consider those finite parts of the fluctuation determinant which are
obtained by omitting the renormalization parts completely.

The numerical computation has been carried out with all the technical
details as described in Ref. \cite{BaaKi}. The results will be discussed
in section 5.
\setcounter{equation}{0}
\section{The Large Mass Approximation}
\par
If the fermion mass $m_F$ becomes much larger than the inverse
size $R^{-1}$ of the background field configuration one expects  the
Dirac operator to  ``seize'' the background field only locally.
Therefore it should be possible to expand the fluctuation determinant
with respect to derivatives which on dimensional grounds
results in an expansion with respect to $1/m_F$. Various
techniques are available. While the heat kernel expansion is
used frequently we use the ordinary Feynman graph or resolvent
expansion here. While heat kernel expansions are often
broken off at a fixed power of the heat kernel time we will
keep track here of {\it all} terms of the leading and next-to-leading
power of the mass $m_F$.

The fermion determinant can be written exactly as
\be
\ln {\cal J} = {\rm Tr} \ln ({\bf 1} + {\bf G}^{(0)}
{\cal V})
\ee
where the free Green function is defined as
\be
{\cal M}{\bf G}^{(0)} = {\bf 1}
\ee

We define the Fourier transform of the potential ${\cal V}$
as
\be
\widetilde{{\cal V}}({\bf q}) = \int d^D x {\cal V}({\bf x}) \exp (- i
{\bf q x}).
\ee
The fluctuation determinant can then be expanded as
\be
\ln {\cal J} = \sum_{n=1}^\infty
{\rm Tr} \frac{(-1)^{n+1}}{n}\int\frac{d^Dp}{(2\pi)^D}
\prod_{j=1}^n \int \frac{d^Dq_j}{(2\pi)^D}
\frac{\widetilde{{\cal V}}({\bf q}_j)}
{({\bf p}+ {\bf Q}_j)^2+m_F^2} (2\pi)^D
\delta^D ({\bf Q}_n)
\ee
where 
\be
{\bf Q}_j = \sum_{l=1}^j {\bf q}_l
\ee
The range of the integration in the variables ${\bf q}_j$ is
determined by the inverse range of the external field configuration
$R^{-1}$. If $m_F \gg R^{-1}$ the scale 
for the integration in ${\bf p}$ is set by the
fermion mass $m_F$. This integration yields therefore a factor
$m_F^{(D-2n)}$. Recalling the form of the potential
\be
{\cal V} = m_F^2 (\phi^2-1) +m_F \gamma \partial \phi
\ee
we see that in leading order the product over the potentials
yields a factor $m_F^{2n}$ so that the leading behaviour of
$\ln {\cal J}$ is $m_F^D$
\footnote{Note that this power counting is different in the
case of chiral mass terms of the form
$m_0 \exp (i \gamma_5 \phi_a \tau_a)$ where only the gradient
term linear in $m_0$ appears in ${\cal V}$.}. 
The next-to-leading order term is
obtained by taking into account the first nonleading terms in the expansion
of the denominators and of the trace of the product over the
potentials. 
Using symmetric integration we have
\bea
\prod_{j=1}^n \frac{1}{({\bf p}+ {\bf Q}_j)^2+m_F^2}
&\simeq&
\frac{1}{({\bf p }^2+m_F^2)^n}\Big[ 1-\sum_{l=1}^n\frac{{\bf Q}_l^2}
{({\bf p}^2+m_F^2)} \nonumber
\\ &+& \frac{4}{D}
\frac{{\bf p}^2}{({\bf p}^2+m_F^2)^2}(\sum_{l>l'}
{\bf Q}_l{\bf Q}_{l'} +\sum_{l=1}^n {\bf Q}_l^2 ) \Big]
\eea
up to terms of order $m_F^{(-2n-4)}$. Similarly we have
\bea
{\rm Tr} \prod_j \widetilde{{\cal V}}({\bf q}_j) &\simeq& \nu_F
m_F^{2n}\Big [\prod_j \widetilde{({\bf \phi}^2-1)} ({\bf q}_j)
 \nonumber \\
&-& \frac{1}{m_F^2}\sum_{l>l'}{\bf q}_l{\bf q}_{l'}\phi({\bf q}_l)
\phi({\bf q}_{l'})\prod_{j \neq l,l'}\widetilde{({\bf\phi}^2-1)}
 ({\bf q}_j)\Big ]
\eea
up to terms of order $m_F^{(2n-4)}$. Here $\nu_F$ is $2$ for
$D=2$ and $4$ in $D=3$ and $D=4$.
Inserting these expansions one obtains after transforming back to
$x$ space
\bea
\ln {\cal J} &=& \nu_F m_F^D \sum_{n=1}^\infty
\frac{(-1)^{n+1}}{n} \int d^Dx
\Big\{ ( I(D,0,n)(\phi^2-1)^n\nonumber \\ & +&
\frac{n(n-1)}{2m_F^2} I(D,0,n)(\phi^2-1)^{n-2} (\partial_\mu \phi)^2
 \\
&-&\frac{2(n+1)n(n-1)}{3m_F^2} I(D,0,n+1) \phi^2
(\phi^2-1)^{n-2}(\partial_\mu \phi)^2    \nonumber \\
&+& \frac{2(n+2)(n+1)n(n-1)}{3Dm_F^2} I(D,1,n+2)\phi^2 (\phi^2-1)^{n-2}
(\partial_\mu \phi)^2\Big\}
 \nonumber
\eea
where 
\be
I(D,m,n)=\int \frac{d^Dl}{(2\pi)^D}\frac{\left({\bf l}^2
\right)^m}{({\bf l}^2+1)^n}
\ee

The sum over $n$ can be performed after evaluating explicitly the
(dimensionless) integrals $I(D,m,n)$ in the various dimensions.
The first few terms are however formal only: the integrals are
divergent. As for the exact computation we have to omit,
for $D=2$ and $D=3$ the order $n=1$
and for $D=4$ the orders $n=1$ and $2$.
We denote the summation over
$n=n_0,..\infty$ by the symbol $\overline{(n_0)}$ as already in the
previous section. 

Writing the result in the form of an effective action we have
for $D=2$
\bea
S_{eff}^{\overline{(2)}}[\phi] = -\frac1{2}
\ln {\cal J}^{\overline{(2)}} &\simeq&
\frac{m_F^2}{4\pi}\int d^2x (1+\phi^2(\ln (\phi^2) -1))
\nonumber \\
&&+\frac{1}{24 \pi}\int d^2x \frac{(\partial_\mu \phi)^2}{\phi^2}
\label{San2}
\eea
for $D=3$
\bea
S_{eff}^{\overline{(2)}}[\phi] = -\frac1{2}
\ln {\cal J}^{\overline{(2)}}
&\simeq&  \frac{m_F^3}{3\pi} \int d^3x (|\phi|^3
-\frac{3}{2} (\phi^2-1) -1) \nonumber \\
&&+\frac{m_F}{12\pi} \int d^3x \frac{(\partial_\mu \phi)^2}{|\phi|}
\label{San3}
\eea
and for $D=4$
\bea
S_{eff}^{\overline{(3)}}[\phi] =\
- \frac1{2} \ln {\cal J}^{\overline{(3)}}&
\simeq& -\frac{m_F^4}{16 \pi^2}
\int d^4x (\phi^4 \ln (\phi^2) -\frac{3}{2}(\phi^2-1)^2 -(\phi^2-1))
\nonumber \\
&&- \frac{m_F^2}{16 \pi^2} \int d^4 x (\partial_\mu \phi)^2
(\ln(\phi^2) -\frac{2}{3} (\phi^2-1))
\label{San4}
\eea
\setcounter{equation}{0}
\section{Results and Conclusions}
We have evaluated the exact fluctuation determinants described in the
previous sections in $D=2,3$ and $4$ for two different typical
profiles, a ``small''and a ``large'' bubble. For convenience we used
an analytical expression for the radial profile
\be
f(r) = (1 + \exp (-\sqrt{r^2+1}+\sqrt{R^2+1}))^{-1}
\label {profdef} 
\ee
This function behaves near $r=0$ as $c_1+ c_2r^2$ in analogy to
the solutions of the typical field equations for the Higgs field,
for $r \to \infty$ it behaves as $1-c_3 \exp(-r)$ as it would 
behave for a Higgs field of mass $1$. 
$R$ determines the point $f(r)=1/2$ and therefore the typical size
of the profile. We have chosen $R=2$ and $R=6$. In the second case
the function $f(r)$ goes to $.006$ at $r=0$ so that the symmetric
Higgs phase is almost reached. This is a situation where breaking
off the expansion in $(\phi^2-1)$ at a finite power of this quantity
would lead to a poor approximation. The profiles are displayed in
Fig. 1.

The results of the exact and approximate computations are
presented in Figs. 2 - 7, both for the small and large sized
profiles and for D=2, 3 and 4. 

The precision of the exact computation can be estimated from the
convergence of the partial wave expansion, from varying the
grid density for the radial variable (we have typically 
$1000$ points in the range $0 < r < 15$) etc. to be better than 
of $10^{-3}$. The agreement with the estimate at large fermion masses
further corroborates this estimate. The precision becomes poorer
if $m_F R$ exceeds a value of $ \sim 12$ since then the asymptotic
regime of the partial wave series, requiring $l \gg m_F R$ is not yet
reached at our maximal value of $l = 25$; while the higher
partial waves {\it are} included using a fit $a/l^3 + b/l^4$ the 
parameters of this fit are then not reliable.

In the case $D=2$ and $R=6$
a bound state eigenvalue approaches $0$ in the S-wave
and  the calculation becomes numerically delicate at $m_F > 3$.
The poor agreement {\it in next-to-leading order} between
asymptotic estimate and exact results which is 
apparent in Fig. 3 is however {\it not} 
due to the numerics which we have checked in all aspects. 
The $\phi^{-2}$ singularity in the kinetic term
of the effective action for $D=2$, Eq. (\ref{San2}), indicates a 
bad convergence of the $1/m_F$ expansion if $\phi$ becomes
small, even if, as on the present case, the integral exists.
In all other cases, for both profiles and all dimensions, the first
two terms of the $1/m_F$ expansion agree very well with the   
exact results for sufficiently large $m_F$. The asymptotic
regime sets in earliest for $D=4$.  

In conclusion we have found in general an excellent agreement between
our exact and approximate evaluations of the fluctuation determinant
of a heavy fermion coupled to an external Higgs field, the
singularities of the gradient terms at $\phi(x)=0$ are 
however potentially dangereous and need further consideration.
The comparison corroborates the precision of the numerical method.
The analytic approximation has the form of an effective action
including all terms with at most two derivatives in the external
field. Its should be interesting to include it into the
action from which the classical profiles are determined.

\section*{Acknowledgements}
We have pleasure in thanking Valeri Kiselev for many useful
discussions. Hiroto So thanks the Japanese Ministry of Education
for a grant. J\"urgen Baacke thanks the DESY, where part of
this work was performed, for hospitality.

\begin{appendix}
\setcounter{equation}{0}
\section{Appendix}
We shall  summarize the notations
and formulas in D=2,3 and 4.
The Euclidean Dirac operator, ${\cal D}$, is defined as

\begin{equation}
{\cal D} = \gamma_{\mu}\partial_{\mu} -m_F \phi (x)
 \nonumber
\end{equation}
\noindent
As gamma matrix convention for $D=2$ we take
\begin{eqnarray*}
\gamma_1&=&\sigma_1  \\
\gamma_2&=&\sigma_2  \\
\gamma_5&=&-\sigma_3  \\
\gamma_{\mu}\gamma_{\nu}&=&\delta_{\mu\nu}
-i\epsilon_{\mu\nu}\gamma_5.
\end{eqnarray*}
\noindent
For $D=3$ \footnote{The three dimensional theory is considered here
as a three dimensional reduction of a four dimensional one as it occurs
in finite temperature quantum field theory} and $D=4$ we use
\begin{equation}
\gamma_{\mu}= \left(\begin{array}{cc}0 & \Sigma_{\mu}\\
\Sigma_{\mu}^{\dagger}&0  \end{array} \right)\nonumber
\nonumber
\end{equation}
\noindent
with
\begin{equation}
\Sigma_{\mu} = \left\{\begin{array}{cl}1_{2\times2}& \mu=4\\
i\sigma_{a} & \mu=a=1\sim 3 \end{array}\right. \nonumber
\end{equation}
\noindent
$\gamma_5$ is then
\begin{equation}
\gamma_5=\left(\begin{array}{cc}-1_{2\times2}&0\\
0&1_{2\times2}\end{array}\right)\nonumber
\end{equation}
\noindent

 This is just the chiral representation.
For a field configuration $\phi$ which depends 
only on the radius, $\phi(x)=f(r)$  the ``squared'' Dirac operator 
${\cal M}$ defined in Eq. (\ref{Mdef}) is given by
\begin{eqnarray*}
{\cal M} &=&  -\partial ^2 + m_F^2 + m_F^2(f(r)^2-1) +
                 m_F f^{\prime}(r) \hat{x}_{\mu}\gamma_{\mu} \\
             &=&  -\displaystyle{\frac{\partial^2}{\partial r^2}
                  -\frac{D-1}{r}\frac{\partial}{\partial r}
                  + \frac{L^2}{r^2}} + m_F^2 + {\cal V}(r)
\end{eqnarray*}
\noindent
where $\hat{x}_{\mu}$ is the unit vector of the coordinate $x_{\mu}$,
potential ${\cal V}$ becomes
\begin{equation}
{\cal V}= m_F^2(f^2-1) + m_F f^{\prime} \hat{x}_{\mu}\gamma_{\mu}
\nonumber
\end{equation}
\noindent
and $L^2/r^2$ is the centrifugal potential. 

We take as a basis for the solutions of the Dirac equation in $D=2$
the spinors 

\begin{equation}
\Psi = \left(\begin{array}{c}\psi_L\\\psi_R\end{array}\right)
= \left(\begin{array}{c} u(r) e^{in\varphi}\\ v(r) e^{i(n+1)\varphi}
\end{array}\right)\nonumber
\end{equation}
\noindent
where the angular momentum $n$ takes the values $n=0,\pm 1, \pm 2..$.
For $D=3$ we use the spinors
\begin{equation}
\Psi = \left(\begin{array}{c}\psi_L\\ \psi_R\end{array}\right)
 = \left(\begin{array}{c}   u(r)\chi\\  i v(r) \hat{x}_{a}
 \sigma_{a}\chi \end{array} \right).
   \nonumber
\end{equation}
For the four dimensional theory we take
\begin{equation}
\Psi = \left(\begin{array}{c}\psi_L\\ \psi_R \end{array} \right)
 = \left( \begin{array}{c}  u(r)\chi\\
 v(r) \hat{x}_{\mu}\Sigma^{\dagger}_{\mu}\chi
\end{array} \right) \nonumber
\end{equation}
\noindent
$\chi$ is a 2-component spinor multiplied by
a spherical harmonic function and
\begin{equation}
L^2 \chi = 4j(j+1)\chi
\nonumber
\end{equation}
\noindent
where $j(=0,1/2,1...)$ is the magnitude of 'angular momentum'
in D=4 and
\begin{equation}
L^2 \chi = j(j+1)\chi
\nonumber
\end{equation}
\noindent
where $j(=0,1,..)$ is the magnitude of angular momentum
in D=3.
 Then the  'squared'  Dirac equations are
\begin{equation}
-u^{\prime\prime}-\displaystyle{\frac{1-2\beta}{r}u^{\prime}
+\frac{\lambda^2-\beta^2}{r^2}}u+m_F^2u + m_F^2(f^2-1)u + m_Ff'v=0
\nonumber
\end{equation}
\begin{equation}
-v^{\prime\prime}-\displaystyle{\frac{1-2\beta}{r}v^{\prime}
+\frac{(\lambda+1)^2-\beta^2}{r^2}}v+m_F^2v + m_F^2(f^2-1)v + m_Ff'u=0
\nonumber
\end{equation}
\noindent
where 
\begin{equation}
\lambda = \left\{
\begin{array}{ll}
  \vert n\vert=0,1,2,.., & {\rm for~ D=2} \\
      j+1/2 =1/2,3/2,5/2,.., & {\rm for~ D=3} \\
      2j=0,1,2,.., & {\rm for~ D=4}
\end{array}\right.
\end{equation}   
and $\beta$ is $\displaystyle{\frac{2-{\rm D}}{2}}$.
In this basis the potential is  simply the $(2 \times 2)$ matrix
\begin{equation}
{\bf V}=
\left(
\begin{array}{cc}
 m_F^2(f^2-1) & m_F f^{\prime}\\
 m_F f^{\prime}& m_F^2(f^2-1)
\end{array}
\right)
\nonumber
\end{equation}
For the free equation($f=1$), the regular solutions are
\begin{eqnarray}
\Psi_1&=&\left( \begin{array}{c}  b_{\lambda}(m_Fr) \\ 0
\end{array}  \right)\\
\Psi_2&=&\left( \begin{array}{c}  0 \\  b_{\lambda+1}(m_Fr)
\end{array} \right)
\nonumber
\end{eqnarray}
where
\begin{equation}
b_{\lambda}(m_Fr) = r^{\beta}I_{\lambda}(m_Fr)
\label{bdef}
\end{equation}   
\noindent
and $I_{\lambda}(m_Fr)$ is the modified Bessel function of
order $\lambda$.
Therefore, the solutions of the 'squared' Dirac equation are
 expressed as
\begin{equation}
\psi_{i \alpha}=(\delta_{i\alpha}+h_{i\alpha}(r))b_{n-1+i}(m_Fr)
\nonumber
\end{equation}
\noindent
where $\alpha$ labels the different  solutions
and $i$ the components of the solution.
The functions $h_{i\alpha}(r)$, are  determined by
\begin{equation}
\Delta_{\lambda,ij} h_{j\alpha} = V_{\lambda,ij}
(\delta_{j\alpha}+h_{j\alpha})
\label{hequation1} \ee
where
\be
\Delta_{\lambda,ij} =
\left[\frac{d^2}{dr^2}  +
\left(\displaystyle{\frac{D-1}{r}+ 2 m_F \frac{b_{\lambda-1+i}'(m_Fr)}
     {b_{\lambda-1+i}(m_Fr)}}\right) \frac{d}{dr}\right] \delta_{ij}
\label{hequation2}
\ee
and
\be
V_{\lambda,ij}=V_{ij}\frac{b_{\lambda-1+j}(m_Fr)}{b_{\lambda-1+i}(m_Fr)}
\label{hequation3}
\end{equation}
\noindent
The degeneracies of the solutions are given by
\begin{equation}
d_{\lambda}({\rm D})=\left\{
\begin{array}{ll}
2,&{\rm for~ D=2}\\
2(2\lambda+1),&{\rm for~ D=3}\\
2 (\lambda+1) (\lambda+2),&{\rm for~ D=4}
\end{array}
\right.
\end{equation}
\noindent
\end{appendix}

\newpage
\section*{Figure Captions}

\begin{description}

\item[Fig. 1] The Higgs field profiles.
We plot the radial profiles given in Eq. (\ref{profdef})
for $R=2$ and $R=6$ as used in our computations.

\item[Fig. 2] The fermionic effective action $S_{eff}^{\overline
{(2)}}/m_F^2$ for $D=2$, $R=2$.
The solid line represents the analytic approximation
of Eq. (\ref{San2}), the numerical exact results are plotted as
points ($\bullet$) connected by straight dotted lines.
The straight dashed line indicates the limit $m_F \to \infty$.

\item[Fig. 3] The same as Fig. 2 for $R=6$.

\item[Fig. 4] The fermionic effective action $S_{eff}^{\overline
{(2)}}/m_F^3$ for $D=3$, $R=2$. Description as for Fig. 2, the analytic 
approximation being given in Eq. (\ref{San3}).

\item[Fig. 5] The same as Fig. 4 for  $R=6$.

\item[Fig. 6] The fermionic effective action $S_{eff}^{\overline
{(3)}}/m_F^4 $ for $D=4$, $R=2$. Description as for Fig. 2, the
analytic approximation being given in Eq. (\ref{San4}).

\item[Fig. 7] The same as Fig. 6 for $R=6$.

\end{description}

\newpage
\pagestyle{empty}
\begin{center}
\textheight29cm
\mbox{\epsfig{file=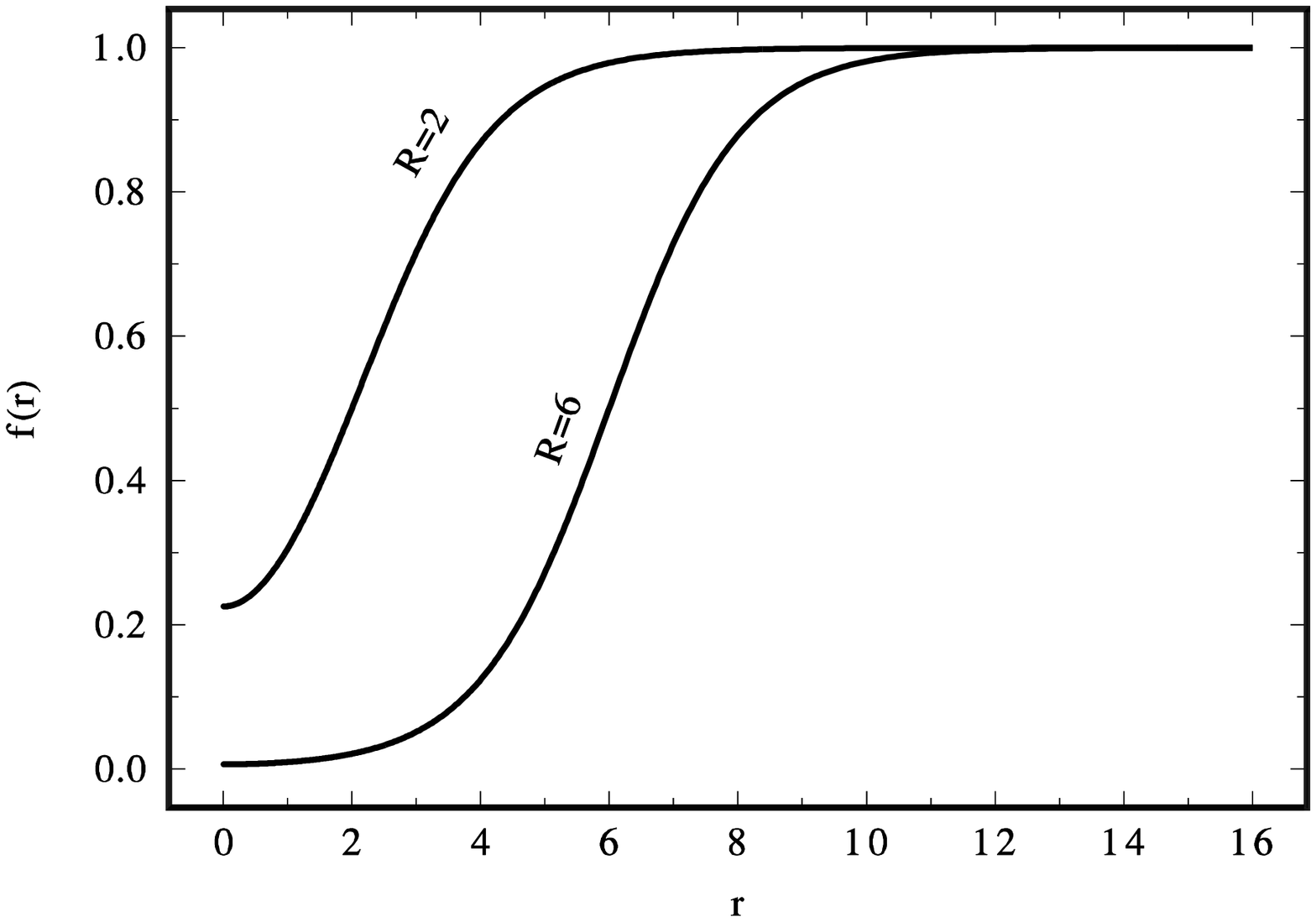,width=14cm,angle=-90}}
\mbox{Figure 1}
\newpage
\mbox{\epsfig{file=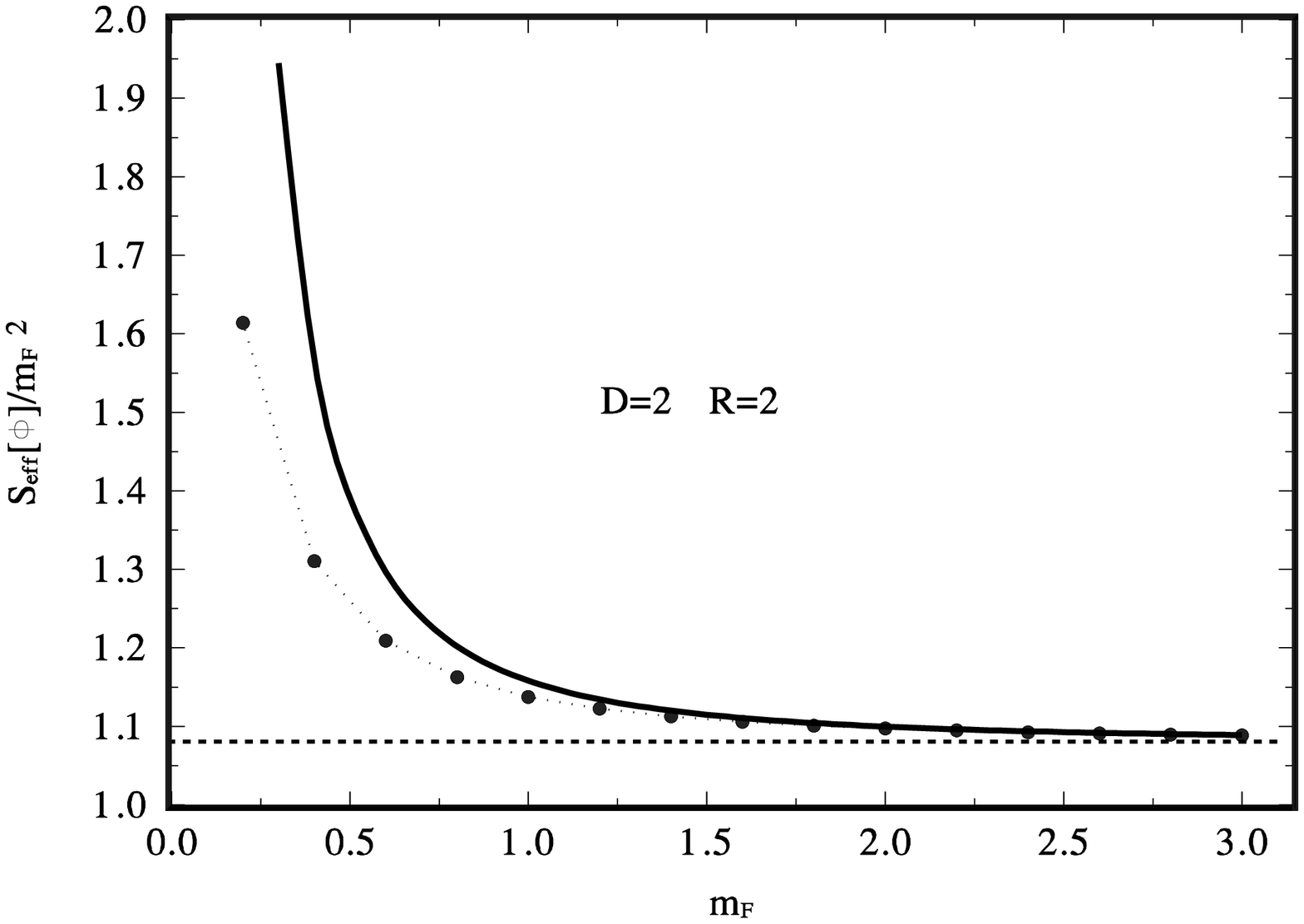,width=14cm,angle=-90}}
\mbox{Figure 2}
\mbox{\epsfig{file=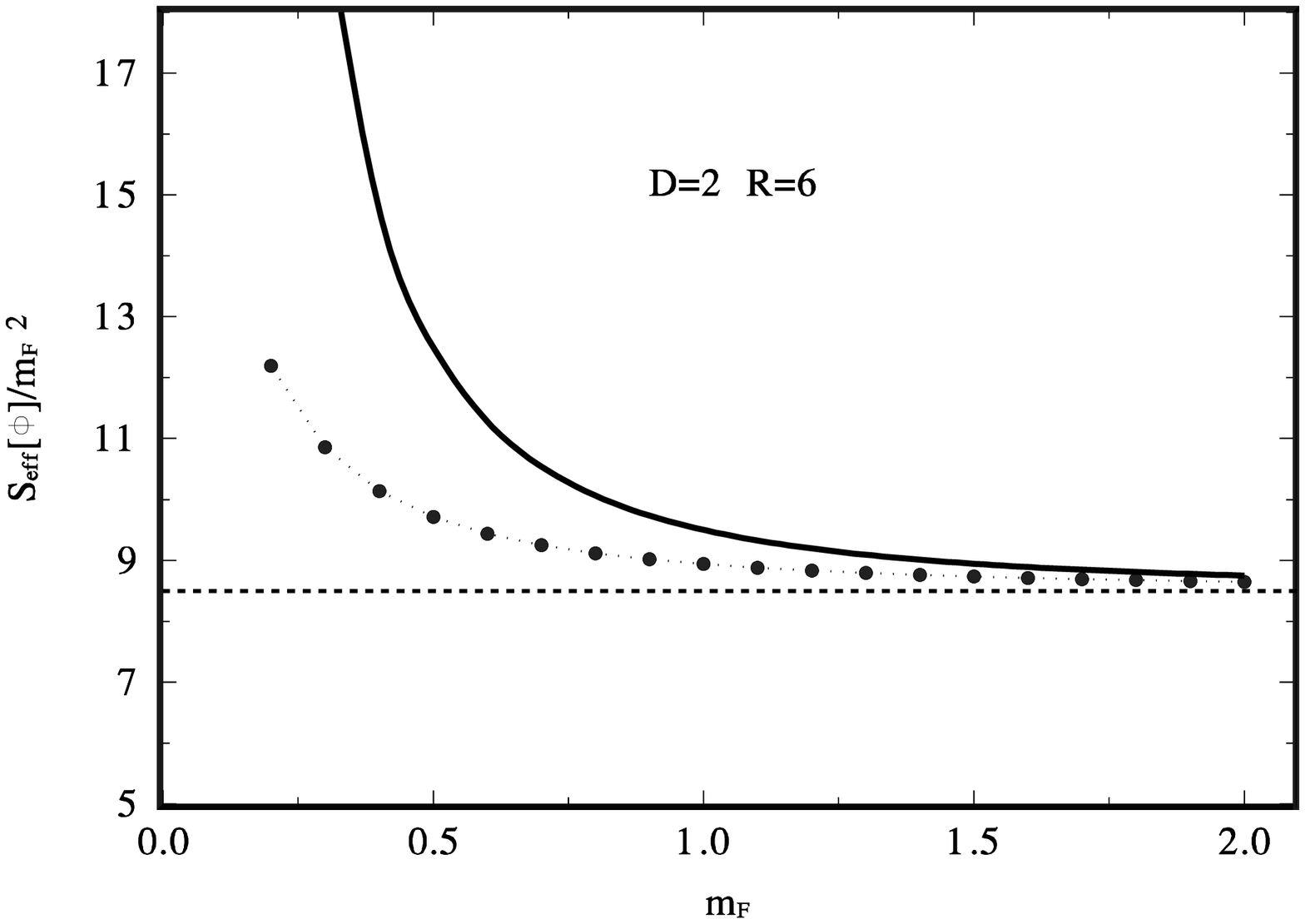,width=14cm,angle=-90}}
\mbox{Figure 3}
\newpage
\mbox{\epsfig{file=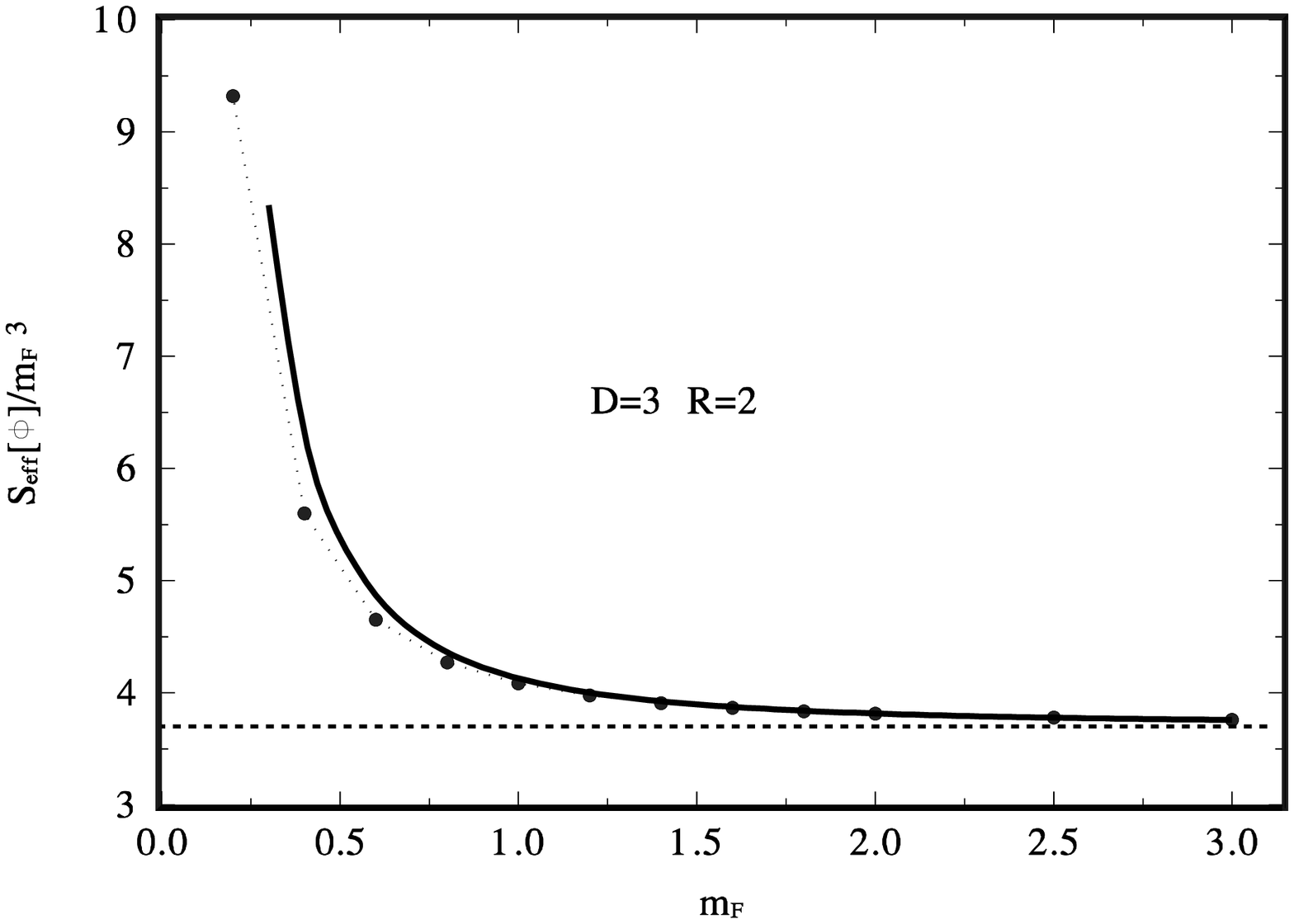,width=14cm,angle=-90}}
\mbox{Figure 4}
\mbox{\epsfig{file=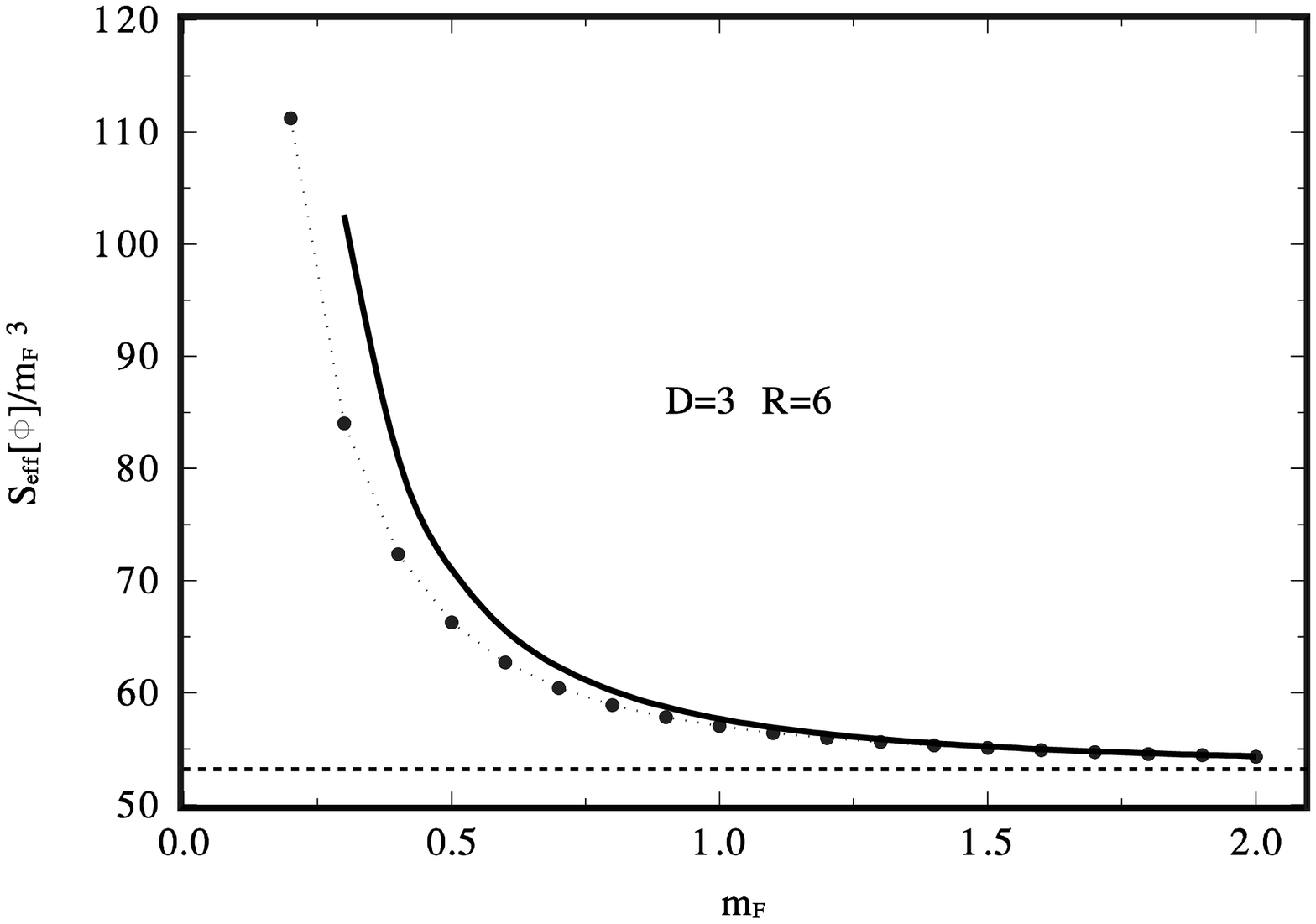,width=14cm,angle=-90}}
\mbox{Figure 5}
\newpage
\mbox{\epsfig{file=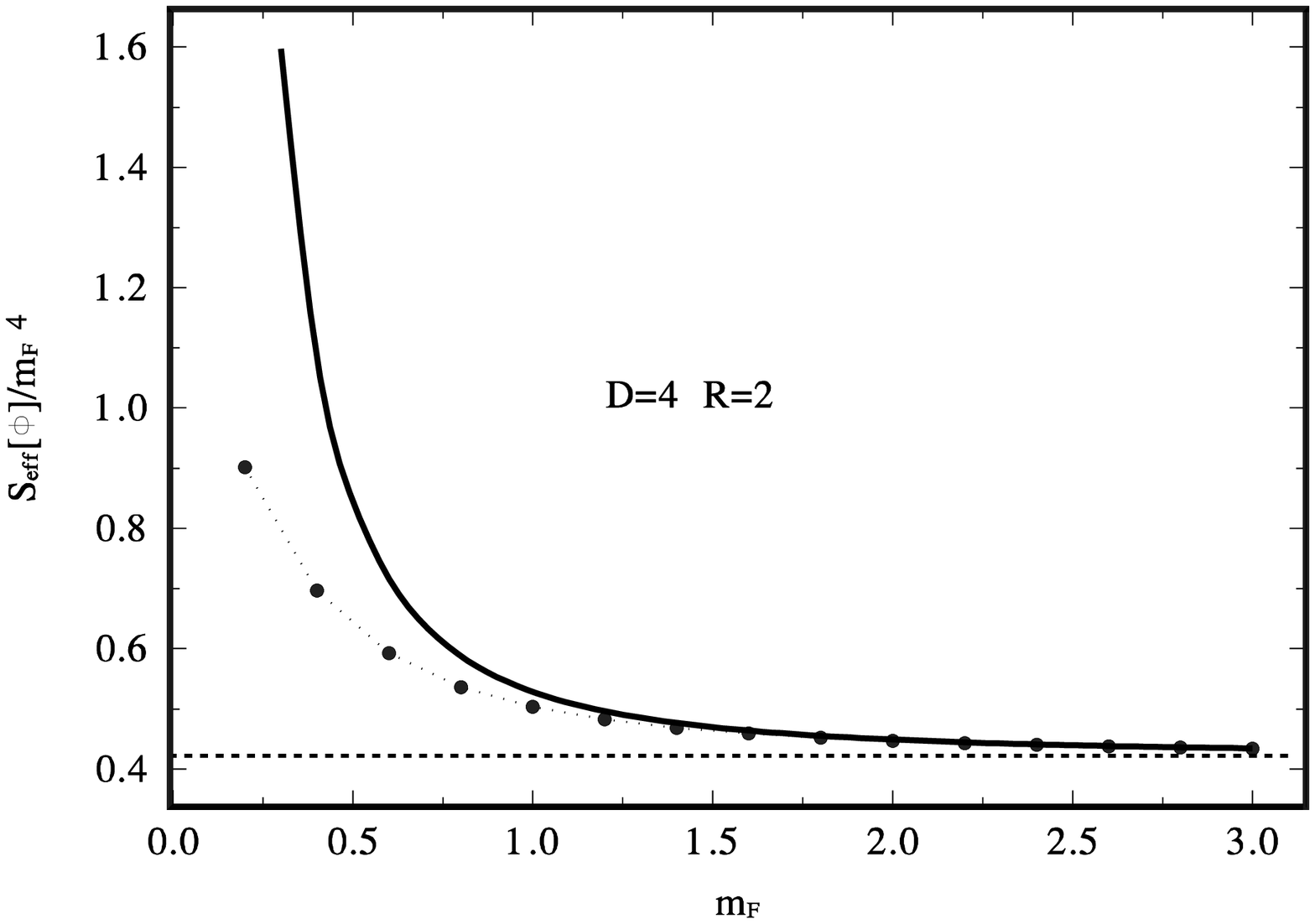,width=14cm,angle=-90}}
\mbox{Figure 6}
\mbox{\epsfig{file=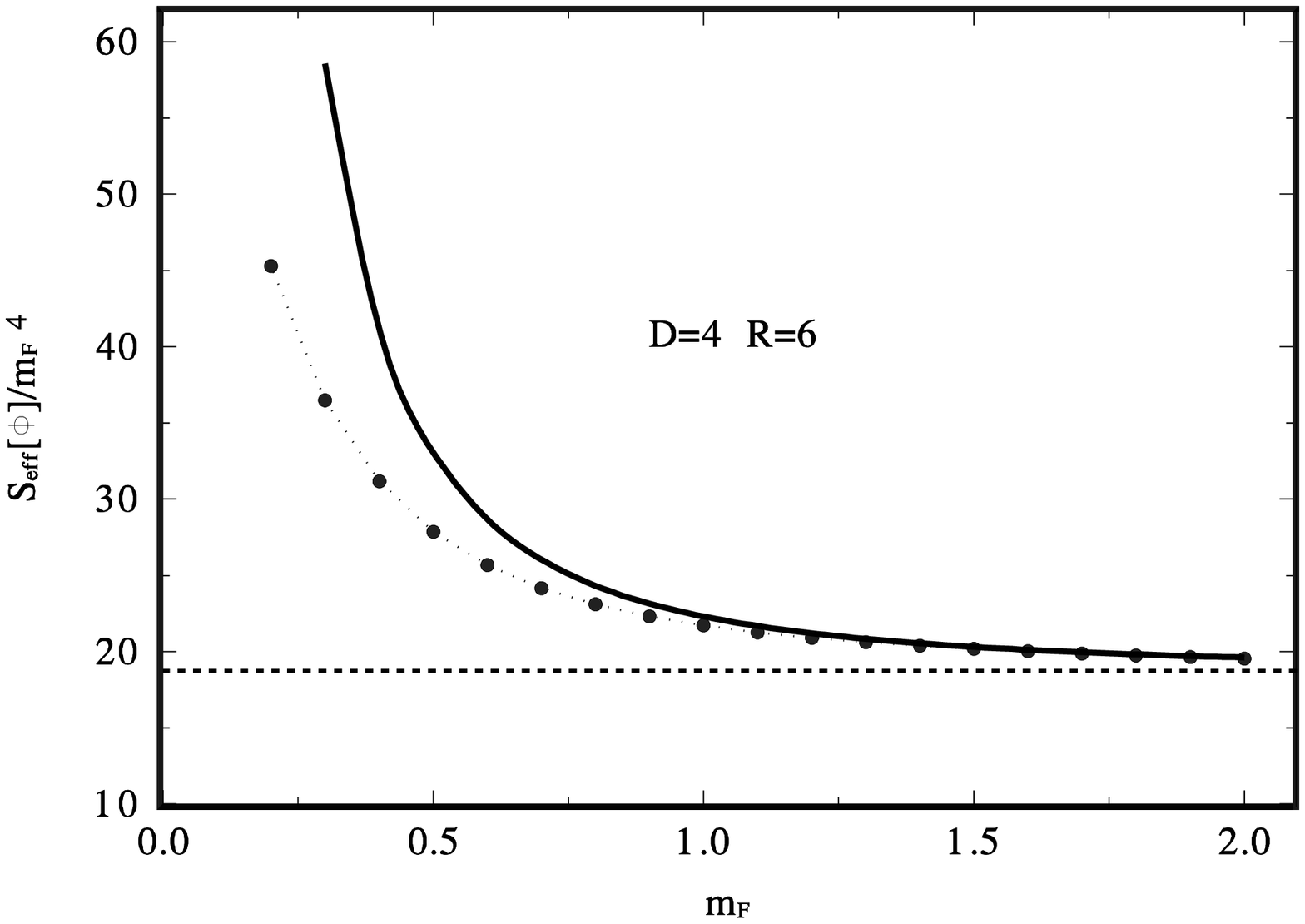,width=14cm,angle=-90}}
\mbox{Figure 7}
\end{center}
\end{document}